
%
%

\magnification \magstep1

\hoffset 1.5truecm
\hsize 16truecm \vsize 23.5truecm
\baselineskip 20pt \parskip5pt
\raggedbottom

\def\coef{{\pi i\over4k}}
\def\half{{1\over2}}


\def\pmbb#1{\setbox0=\hbox{I}\setbox1=\hbox{#1}
\dimen0=.5\wd0\advance\dimen0 by .05\wd1
\box0\kern-\dimen0\box1}

\def\B{\pmbb{B}}
\def\E{\pmbb{E}}
\def\I{\pmbb{I}}


\def\bigplusskein{
\setbox0=\hbox{
\setbox1=\hbox{\bigg/}
\setbox2=\hbox{$\backslash$}
\hbox to -\wd1{} \copy1
\kern-\wd1\raise.5\ht1\copy2
\lower\dp1\hbox{\raise\dp2\copy2}
}\copy0}

\def\bigminusskein{
\setbox0=\hbox{
\setbox1=\hbox{$\bigg\backslash$}
\setbox2=\hbox{/}
\hbox to -\wd1{} \copy1
\kern-\wd1\lower.5\ht1\copy2
\raise\dp1\hbox{\lower\dp2\copy2}
}\copy0}

\def\bigzeroskein{
\;\bigg\vert\;\bigg\vert\;}

\def\biginftyskein{\;
{\bigcup\atop\bigcap}\;}

\def\bignullskein#1{\;\bigg\vert #1\;}

{\parindent=0pt

\rightline{hep-th/9302021}

\vfill

{\bf SU(2) and the Kauffman bracket}

B Broda\footnote{\dag}{Permanent address: Department
of Theoretical Physics, University of \L\'od\'z, Pomorska
149/153, PL--90-236 \L\'od\'z, Poland; e-mail:
ptbb@ibm.rz.tu-clausthal.de}

Institute for Mathematical Physics, Technical University of
Clausthal, Leibnizstra\ss e 10, D-W--3392
Clausthal-Zellerfeld, Federal Republic of Germany

\vfill

A direct relationship between the (non-quantum) group SU(2)
and the Kauffman bracket in the framework of Chern-Simons
theory is explicitly shown.
}
\vfill\eject

In his seminal paper on quantum field theory and the Jones
polynomial [1], Edward Witten proposed, in the framework of
Chern-Simons theory based on the group SU(2), a new
approach to invariants of knots and links. In this letter,
we would like to show that the mathematical object directly
related to the fundamental representation of SU(2) is the
Kauffman bracket [2]. Strictly speaking, we will show that
SU(2) Chern-Simons theory at the level $k$, with line
observables defined in the fundamental representation,
directly corresponds to the (one-variable/specialization of
the) Kauffman bracket (regular isotopy invariant of
unoriented links) with the parameter $A=\exp(-\pi i/4k)$.

The ``half-monodromy'' or (quasi-)braiding matrix derived
from the Chern-Simons theory is of the form [3--4]
$$
\B=\exp\left(-{\pi i\over k}{\bf t\/}^{\otimes2}\right),
\qquad
k\in{\bf Z\/}^\pm,
\eqno(1)
$$
where $\bf t$ is a generator of a compact semi-simple Lie
group $G$. Putting for $G=\rm SU(2)$ $t^a=\half\sigma^a$,
$a=1,2,3$ ($\sigma\/$'s are the Pauli matrices), and using
the Fierz identity, we obtain
$$
\eqalignno{
\B&=\exp\left(-\coef
\sum_{a=1}^3\sigma_{ij}^a\otimes\sigma_{kl}^a\right)\cr
&=\exp\left[\coef(\I-2\E)\right]\cr
&=\exp\left(\coef\right)\left(\I\cos{\pi\over2k}
-i\E\sin{\pi\over 2k}\right),
&(2)\cr}
$$
with
$$
\I\equiv\delta_{ij}\delta_{kl},
\qquad
\E\equiv\delta_{il}\delta_{kj}.
\eqno(3)
$$
The correspondence
$$
\eqalignno{
\B&\longleftrightarrow\bigplusskein\cr
\B^{-1}&\longleftrightarrow\bigminusskein\cr
\I&\longleftrightarrow\bigzeroskein&(4)\cr}
$$
yields the following skein relation
$$
A\left<\bigplusskein\right>-A^{-1}\left<\bigminusskein\right>
=(A^2-A^{-2})\left<\bigzeroskein\right>,
\eqno(5)
$$
where
$$
A=\exp\left(-\coef\right),
\eqno(6)
$$
and ``$\left<\;\right>$'' denotes the normalized
quantum-field-theory expectation value with respect to the
Chern-Simons ``measure''. Rotating the graphs in (5), we obtain
a new equivalent skein relation
$$
A\left<\bigminusskein\right>-A^{-1}\left<\bigplusskein\right>
=\left(A^2-A^{-2}\right)\left<\biginftyskein\right>.
\eqno(7)
$$
Combining (5) and (7) produces
$$
\left<\bigminusskein\right>=A\left<\biginftyskein\right>
+A^{-1}\left<\bigzeroskein\right>.
\eqno(8)
$$
All the lines entering our graphs should be unoriented. It
follows from the fact that the fundamental representation
of the group SU(2) is non-complex (pseudo-real) [5], and
the expectation values of the line observables in the
fundamental representation have to be invariant with
respect to the reversing of orientation [4].

To compute the dependence of a line on a framing one should
contract two indices in the exponent of $\B$ (say, $j$ and
$k$) yielding
$$
\exp\left(-{\textstyle3\pi i\over4k}\right)=A^3.
\eqno(9)
$$
Thus,
$$
\left<\bignullskein{\pm1}\right>
=-A^{\pm3}\left<\bignullskein{0}\right>,
\eqno(10)
$$
where the minus sign follows from the pseudo-reality of the
fundamental representation of SU(2), and the integers
mean the framing. Closing in (8) the left legs of all the
(three) diagrams with arcs, as well as the right ones, and
applying (10), we obtain
$$
\left<\bigcirc\right>=-A^2-A^{-2}.
\eqno(11)
$$
In (11), we have used the property of locality of
Chern-Simons theory, which can be expressed as
$$
\left<L_1\sqcup L_2\right>=\left<L_1\right>\left<L_2\right>,
\eqno(12)
$$
where the symbol ``$\sqcup$'' denotes a distant union of
links (links separated with a two-sphere).

Collecting equations (8), (11) and (12) we can write down
the full set of the axioms of the (one-variable) Kauffman
bracket:

\noindent
(i)
$$
\left<\emptyset\right>=1
\eqno(13c)
$$
\noindent
(ii)
$$
\left<\bigcirc\sqcup L\right>=\left(-A^2-A^{-2}\right)\left<L\right>
\eqno(13b)
$$
\noindent
(iii)
$$
\left<\bigminusskein\right>=A\left<\biginftyskein\right>
+A^{-1}\left<\bigzeroskein\right>,
\eqno(13a)
$$
where $A=\exp(-\pi i/4k)$.

\bigskip
\noindent
{\bf Acknowledgments}

The author is indebted to Professor H.~D.~Doebner for his
kind hospitality in Clausthal. The work was supported by
the Alexander von Humboldt Foundation and the KBN grant
202189101.

\vfill\eject
\noindent\frenchspacing\parskip=0pt
{\bf References}

\item{[1]} Witten E 1989 {\it Commun. Math. Phys.}
{\bf 121} 351

\item{[2]} Kauffman L H 1987 {\it Topology} {\bf 26} 395

\item{} Kauffman L H 1991 {\it Knots and Physics} (World
Scientific)

\item{[3]} Broda B 1990 {\it Mod. Phys. Lett.} A {\bf 5} 2747

\item{[4]} Guadagnini E 1992 {\it Int. J. Mod. Phys.} A {\bf 7}
877

\item{[5]} O'Raifeartaigh L 1988 {\it Group Structure of
Group Theories} (Cambridge: CUP) p~57

\bye